\documentclass[12pt]{article}
\usepackage{latexsym}
\usepackage{amsmath,,calrsfs}
\usepackage{amsfonts}
\usepackage{amssymb}
\usepackage{amscd}
\usepackage{fancybox}
\usepackage{cite}
\usepackage{amsmath,amsfonts,amsbsy}
\usepackage{pstricks,pst-node}
\usepackage[small,bf,hang]{caption2}
\usepackage{psfrag}

\usepackage{float}

\psset{unit=1.3cm,linewidth=.5pt,radius=.2}  

\usepackage{enumitem}                       
\usepackage{multirow}                     
\usepackage{float}                          
\usepackage{lscape}                         
\usepackage{bm}

\def\hybrid{\topmargin -20pt    \oddsidemargin 0pt
        \headheight 0pt \headsep 0pt
        \textwidth 6.25in       
        \textheight 9.5in       
        \marginparwidth .875in
        \parskip 5pt plus 1pt   \jot = 1.5ex}

\hybrid

\newcommand{\beq}{\begin{equation}}
\newcommand{\eeq}{\end{equation}}
\newcommand{\bi}{\begin{itemize}}
\newcommand{\ei}{\end{itemize}}
\newcommand{\bt}{\begin{tabular}}
\newcommand{\et}{\end{tabular}}
\newcommand{\bc}{\begin{center}}
\newcommand{\ec}{\end{center}}

\newcommand{\be}{\begin{equation}}
\newcommand{\ee}{\end{equation}}
\newcommand{\bea}{\begin{eqnarray}}
\newcommand{\eea}{\end{eqnarray}}
\newcommand{\ba}{\begin{array}}
\newcommand{\ea}{\end{array}}

\def\bbox{{\,\lower0.9pt\vbox{\hrule \hbox{\vrule height 0.2 cm
\hskip 0.2 cm \vrule height 0.2 cm}\hrule}\,}}
\newcommand{\dsl}{\pa \kern-0.5em /}

\begin{document}

\begin{titlepage}
\begin{center}

{\Large \bf On the entropy associated with the interior of a black hole}

\vskip 1cm

{\bf Baocheng Zhang}

\vskip 25pt

{\em School of Mathematics and Physics, China University of Geosciences, Wuhan
430074, China
\vskip 5pt }

{email: {\tt zhangbc.zhang@yahoo.com}} \\

\vskip 15pt

\end{center}

\vskip 0.5cm

\begin{center} {\bf ABSTRACT}\\[3ex]\end{center}

The investigation about the volume of a black hole is closely related to the
quantum nature of the black hole. The entropy is a significant concept for
this. A recent work by Majhi and Samanta [Phys. Lett. B 770 (2017) 314] after
us presented a similar conclusion that the entropy associated with the volume
is proportional to the surface area of the black hole, but the proportionality
coefficient is different from our earlier result. In this paper, we clarify
the difference and show that their calculation is unrelated to the interior of
the black hole.

\end{titlepage}

\newpage

\section{Introduction}

Recently, Christodoulou and Rovelli (CR) suggested a definition of
\textquotedblleft volume\textquotedblright\ for a collapsed black hole
\cite{cr15}, which bridges between the interior of black holes and
thermodynamics \cite{bz15}. The calculation of entropy associated with the
volume is vital for building this connection. We firstly calculated such
entropy statistically and found that the entropy is proportional to the
surface area of the black hole \cite{bz15}. In particular, the thermodynamics
corresponding to the volume entropy is balanced by the contribution from the
volume. Thus, the first law of black hole thermodynamics is not modified,
unlike the situation in which the cosmological constant is considered
\cite{cgp11,bpd11,lgz17}. We also found that the semi-classical result is not
enough to interpret the black hole entropy statistically, which is the
motivation for us to continue the calculation at Planck level with the aid of
spatial noncommutativity. Thus, we found an example that a black hole has
infinite volume but wrapped by finite surface \cite{zy17}, which supports the
conjecture \cite{jmr05,mkp06} that the black hole entropy can be independent
of the black hole interior.

While Majhi and Samanta reinvestigated this volume entropy after us, they got
a different result and pointed out that the difference was due to our
\textquotedblleft improper\textquotedblright\ treatments \cite{ms17}. We have
to say that all our \textquotedblleft improper\textquotedblright%
\ treatments\ were also adopted in the calculation made by Majhi and Samanta.
Firstly, their vital result is the expression of the energy, i.e. Eq. (39) in
their paper \cite{ms17}, which derives from the relativistic dispersion
relation or the primary constraint in their paper. In fact, this constraint
condition is consistent with WKB approximation. In this paper, we will present
this point and discuss the feasibility of this approximation although we had
discussed this in our earlier paper \cite{bz15}. The second point is that when
we calculated the entropy from the free energy, we didn't make the derivative
with respect to the inverse temperature for the CR volume. Actually, we once
calculated this term and found that it gives a very small constant, so in our
earlier paper, we wrote such words \textquotedblleft Temporarily ignoring the
exotic feature of the CR volume\textquotedblright\ when we calculated the
entropy statistically from the free energy. In this paper, we will present
this point in detail. The final point pointed out by Majhi and Samanta is that
the Stefan-Boltzmann law is not valid in the final stage of black hole
evaporation, which had been discussed in detail in our earlier paper.
Actually, this is the reason that we introduce the spatial noncommutativity to
continue to finish the calculation, i.e. see our paper \cite{zy17}. Similarly,
the work made by Majhi and Samanta had the same question and cannot extend to
the final stage of evaporation.

So, since their treatment is consistent with ours, why does the expression of
the entropy associated with the volume obtained by Majhi and Samanta is
different from ours? Because their calculation is not related to the interior
(volume) of the Schwarzschild black hole. This is the reason that we decide to
write this paper, not only for clarifying our calculation, but also for
clarifying the fact that the entropy obtained in Ref. \cite{ms17} is not that
corresponding to the interior of a Schwarzschild black hole as they claimed.
Throughout this paper, we use units with $G=c=\hbar=k_{B}=1$.

\section{CR volume}

We start with the CR definition \cite{cr15} for a collapsed black hole, which
is based on finding the maximal space-like hypersurface $\Sigma$ bounded by a
given surface $S$. For example, given a 2-dimensional sphere $S$ in flat
spacetime by $t=$ $0$ and $r^{2}=x^{2}+y^{2}+z^{2}=R^{2}$, the volume
surrounded by the hypersurface $t=t(r)$ becomes
\begin{equation}
V=\int_{0}^{R}4\pi r^{2}\sqrt{1-\left(  \frac{dt}{dr}\right)  ^{2}}\,\,dr,
\label{uv}%
\end{equation}
which gives $V={4\pi R^{3}}/{3}$ for a sphere by maximizing the hypersurface
with $t=$ constant. A similar discussion can be applied to collapsed matter
described by the Eddington-Finkelstein coordinates
\begin{equation}
ds^{2}=-f(r)dv^{2}+2dvdr+r^{2}d\varphi^{2}+r^{2}\sin^{2}\varphi d\phi^{2},
\label{ef}%
\end{equation}
where $f(r)=1-{2M}/{r}$ and the advanced time $v=t+\int{dr}/{f(r)}%
=t+r+2M\ln\left\vert r-2M\right\vert $. The collapsed matter forms a
Schwarzschild black hole in the end with its event horizon at $r=2M$ serving
as the required surface that bounds many space-like hypersurfaces. Any one
hypersurface $\Sigma$ can be coordinatized by $\lambda,\varphi,\phi$
\cite{cr15}, and the line element of the induced metric on it can be expressed
as
\begin{equation}
ds_{\Sigma}^{2}=\left[  -f(r)\dot{v}^{2}+2\dot{v}\dot{r}\right]  d\lambda
^{2}+r^{2}d\varphi^{2}+r^{2}\sin^{2}\varphi\,d\phi^{2} \label{sv}%
\end{equation}
where the dot represents a partial derivative with regard to the parameter
$\lambda$, and $-f(r)\dot{v}^{2}+2\dot{v}\dot{r}>0$ for a spacelike
hypersurface. Its volume takes the usual form
\begin{align}
V_{\Sigma}  &  =\int d\lambda d\varphi d\phi\sqrt{r^{4}\left[  -f(r)\dot
{v}^{2}+2\dot{v}\dot{r}\right]  \sin^{2}\varphi}\,\nonumber\\
&  =4\pi\int d\lambda\sqrt{r^{4}\left[  -f(r)\dot{v}^{2}+2\dot{v}\dot
{r}\right]  }\,. \label{vi}%
\end{align}
The maximization is obtained at $r={3}M/{2}$ by the method of auxiliary
manifold \cite{cr15} or the method of maximal slicing of mathematical
relativity \cite{bz15}. Carrying out the integration with maximization
condition \cite{cr15,bj15,yco15,yco152}, it is found the CR volume at late
time
\begin{equation}
V_{\mathrm{CR}}\sim3\sqrt{3}\pi M^{2}v. \label{mv}%
\end{equation}
This interesting result shows that volume is determined by advanced time,
which depends on the future behavior, e.g. evaporation \cite{swh74}.

\section{Entropy associated with CR volume}

Based on the above statement about CR volume, we calculated the statistical
entropy in this volume with the phase-space \cite{quan} labeled by position
$\{\lambda,\varphi,\phi\}$ and momentum $\{p_{\lambda},p_{\varphi},p_{\phi}%
\}$. The total number of quantum states arises from integrating ${d\lambda
d\varphi d\phi dp_{\lambda}dp_{\varphi}dp_{\phi}}/{\left(  2\pi\right)  ^{3}}$
over the complete phase space. The integration is carried out by considering a
massless scalar field $\Phi$ in the spacetime with the metric
\begin{equation}
ds^{2}=-dT^{2}+[-f(r)\dot{v}^{2}+2\dot{v}\dot{r}]d\lambda^{2}+r^{2}%
d\varphi^{2}+r^{2}\sin^{2}\varphi d\phi^{2}, \label{zm}%
\end{equation}
which comes from Eq. (\ref{ef}) by the transformation $dv=\frac{-1}{\sqrt{-f}%
}\,dT+d\lambda$ and $dr=\sqrt{-f}\,dT$.

But Majhi and Samanta took the phase space labeled by $\{r,\nu\}$ and momentum
$\{p_{r},p_{\upsilon}\}$ and considered the particle moving in such a
background metric%
\begin{equation}
ds_{ansatz}^{2}=-dt^{2}+r^{4}\left(  -f(r)dv^{2}+2dvdr\right)  . \label{msm}%
\end{equation}
They claimed that in such ansatz, the non-static property could be avoided.

The first and most important problem which requires to be clarified: is the
metric given in Eq. (\ref{msm}) related to the interior of the Schwarzschild
black hole? Obviously, it is unrelated. It is noted that the effective metric
$ds_{eff}^{2}=r^{4}\left(  -f(r)dv^{2}+2dvdr\right)  $ is only an auxiliary
manifold taken for the calculation of the maximization of the volume
(\ref{vi}) \cite{cr15}. So it is not the maximal hypersurface. The maximal
hypersurface is expressed in Eq. (\ref{sv}) with $r={3}M/{2}$. As for the
sphere in flat spacetime, its volume can be defined according to the
maximization of Eq. (\ref{uv}). When the motion of particles needs to be
discussed in this sphere, can it be taken in such background $ds_{eff}%
^{2}=r^{4}\left(  -dt^{2}+dr^{2}\right)  $? Of course, it can not!

Then, is the metric (\ref{msm}) static as claimed by Majhi and Samanta? It is
difficult to give a positive answer. In the statistical calculation for the
modes of the scalar field, the interior spacetime has to be used. The metric
(\ref{zm}) is obviously equivalent to the interior metric (\ref{ef}), but the
metric (\ref{msm}) is not. If ones want to used the metric (\ref{msm}) to
represent the interior of the black hole, a transformation has to be made to
connect the metric (\ref{msm}) with Eddington-Finkelstein coordinates, which
will be found that the coordinate $t$ will not be independent on the
coordinates $r$ and $v$. Thus, the background given by the ansatz is not
static. In general, the non-static characteristics is essential for the black
hole interior, and cannot be taken out by the transformation of coordinates.
So, either the metric (\ref{msm}) is not related to the interior of the black
hole, which is the case made in Ref. \cite{ms17}, or it is not static, which
will lead to the failure of the Hamiltonian analysis. Moreover, the metric
(\ref{msm}) is strange, because both the coordinates $t$ and $v$ are both timelike.

The second problem is that the azimuthal coordinates and their conjugated
momenta were not considered in phase space by Majhi and Samanta. This is
improper, since the scalar field is moving in the spacetime including the
azimuthal directions. If the azimuthal degrees of freedom is ignored in the
construction of phase space, to say the least, how is the integral result
4$\pi$ in the expression (\ref{vi}) of CR volume incorporated into the entropy
associated with the volume? In particular, the spherical symmetry is
represented by these azimuthal coordinates, which will constrain the possible
modes for the scalar field and lead to a smaller statistical result than
having no such consideration. This is the reason why our result is smaller
than that obtained by Majhi and Samanta.

Now it is clear that the calculation made by Majhi and Samanta is not related
to the interior of the black hole and is incomplete even in their ansatz. In
what follows, we will clarify our calculation, and at first interpret the
feasibility of WKB approximation. The main reason to make the approximation is
that our calculation is to be carried out at $v>>M$ and $r={3}M/{2}$, which is
essentially the hypersurface of $T\!\!=$\thinspace constant and is unaffected
by the non-static nature of the metric. Actually, the WKB approximation is
held only if the evolution of spacetime is slow enough near the maximal
hypersurface. It is better to understand this point by maximal slicing
\cite{ewt73,cim11,eg07} in which the evolution avoids the singularity, but the
physical process should be equivalent since in essence the complete theory
should slice-independent. As well-known, in that case, at the final stage of
the hypersurfaces' evolution, a phenomenon called \textquotedblleft collapse
of the lapse\textquotedblright\ happens, which means that near the maximal
hypersurface, the proper time between two neighboring hypersurfaces tends to
zero and no evolution happens there. Thus, the hypersurface is nearly static
and nearly no lapse into the next one. So slow evolution is just the
requirement by the WKB approximation.

Under WKB approximation, the Klein-Gordon equation in curved spacetime gives
the energy expression for the scalar field,%
\begin{equation}
E^{2}-\frac{1}{-f(r)\dot{v}^{2}+2\dot{v}\dot{r}}p_{\lambda}^{2}-\frac{1}%
{r^{2}}p_{\varphi}^{2}-\frac{1}{r^{2}\sin^{2}\varphi}p_{\phi}^{2}=0,
\label{es}%
\end{equation}
which is just the primary constraint $p^{2}=0$ for massless field obtained
using the Hamiltonian analysis of Ref. \cite{ms17}, but the Eq. (\ref{zm})
while not Eq. (\ref{msm}) is taken as the required spacetime metric. With the
expression of energy, the number of quantum states and free energy can be
calculated with the standard statistical method \cite{quan}. So, the entropy
associated with the CR volume is expressed as%
\begin{equation}
S_{\mathrm{CR}}=\beta^{2}\frac{\partial F}{\partial\beta}=\frac{\pi
^{2}V_{\mathrm{CR}}}{45\beta^{3}}-\frac{\pi^{2}}{180\beta^{4}}\frac{\partial
V_{\mathrm{CR}}}{\partial\beta}.
\end{equation}
where the free energy $F=-\frac{\pi^{2}V_{\mathrm{CR}}}{180\beta^{4}}$
\cite{bz15}. With the CR volume in Eq. (\ref{mv}) and $\beta=T^{-1}=8\pi M$,
the second term is constant and the value is approximately $10^{-7}$ which is
less and less than the first term. This is the reason why we didn't include
the second term in our earlier paper. Here, $\beta=T^{-1}=8\pi M$ is an
important assumption that is guaranteed by the condition that the system is in
equilibrium. It is consistent with our calculation with the Stefan-Boltzmann
law. This assumption could also be confirmed and understood by the first law
of black hole thermodynamics which was given in our earlier paper \cite{bz15}.
As presented above, the calculation of CR volume is made in the case of $v>>M$
in which the black hole has formed by the collapse of the matter and is
static\footnote{More rigorously, it should be quasi-static since the black hole
emits Hawking radiation and is gradually evaporating.}
for external observers. Thus, ones can construct the first law of
thermodynamics, which defines an equilibrium state that means that the inside
and outside of the black hole separated by the horizon is in equilibrium. In
particular, it is noted that the interior of the black hole is static when our
statistical calculation is made at the hypersurface of $T\!\!=$\thinspace
constant. The avoidance of non-static nature make the definition of the
temperature in the interior feasible. Since the exterior temperature is the
Hawking temperature for the observer at infinity, the interior temperature can
also be taken as the Hawking temperature for the same observer by the
requirement of equilibrium. In particular, such assumption remains the
validity of the first law and provides an interpretation for the vacuum
pressure at horizon from the perspective of thermodynamics which will be
discussed in the next section.

So the volume entropy is found
\begin{equation}
S_{\mathrm{CR}}\sim\frac{3\sqrt{3}\,}{45\times8^{3}}M^{2}=\frac{3\sqrt{3}%
\,}{90\times8^{4}\pi}A, \label{scr5}%
\end{equation}
where $A=16\pi M^{2}$ is the event horizon area for a Schwarzschild black hole.

\section{Thermodynamic significance}

The entropy $S_{\mathrm{CR}}$ Eq. (\ref{scr5}) associated with $V_{\mathrm{CR}%
}$ remains insufficient to interpret $S_{H}$ because the prefactor in front of
$A$ is much smaller than ${1}/{4}$, but the thermodynamics associated with the
volume has to be considered in the first law \cite{bch73,rb02}. In the
original calculation of Hawking \cite{swh74}, the concept of particles is used
in the asymptotically flat region far away from a black hole where particles
can be unambiguously defined. While the particle flux carries away positive
energy, an accompany flux of negative energy ones falls into the hole across
the horizon, which can only be understood by the zero point fluctuations of
the local energy density in a quantum theory. This phenomena is called vacuum
polarization \cite{fn98} which causes a quantum pressure $P=1/{\left(
90\times8^{4}\pi^{2}{M^{4}}\right)  }$ at the horizon
\cite{wgu76,pc80,dnp82,te83}. It gives $PdV_{\mathrm{CR}}\sim10^{-5}dM$ that
nicely balances out $TdS_{\mathrm{CR}}\sim10^{-5}dM$ on the level of $10^{-5}$
and supports the introduction of the volume $V_{\mathrm{CR}}$
thermodynamically interpretable together with the quantum pressure of vacuum
polarization. It presents not only the connection of volume with quantum
properties of the gravitational field, but also the robustness of the first
law of black hole thermodynamics. At the same time, this in turn shows that
little information can be leaked out from radiation emission except those
about the vacuum polarization and thus indirectly confirms the thermal result
of Hawking radiations. But the result of Majhi and Samanta indicated that the
first law has to be broken, since the terms from the volume in their
expression cannot be balanced out from the two sides of the equation for the
first law. Thus, from this perspective of semi-classical calculation, their
result is also not credible.

In conclusion, our result about the entropy associated with the CR volume is
self-consistent in the semiclassical region, and the reason of the difference
from the calculation made by Majhi and Samanta is that they choose an
erroneous ansatz for the spacetime metric, in which they calculated an
entropy. But its connection with the interior of the Schwarzschild black hole
is unjustified.

\bigskip

We thank Dr. Wu for reminding us of the paper by Majhi and Samanta. This work
is supported by the NSFC with Grant No. 11374330, No. 91636213 and No. 11654001.

\end{document}